\documentclass[journal]{IEEEtran}

\usepackage{cite}
\ifCLASSINFOpdf
\usepackage[pdftex]{graphicx}
\else
   \usepackage[dvips]{graphicx}
\fi
\usepackage[cmex10]{amsmath}
\begin{document}
%

\pagestyle{empty}
\title{Leakage Currents and Capacitances of Thick CZT Detectors}

\author{Alfred~Garson~III$^{1}$, Qiang~Li$^{1}$        
        Ira~V.~Jung$^{2}$, Paul~Dowkontt$^{1}$, Richard~Bose$^{1}$, Garry~Simburger$^{1}$, and~Henric~Krawczynski$^{1}$%
\thanks{$^{1}$Washington University in St. Louis, 1 Brookings Drive, CB 
    1105, St.\ Louis, Mo, 63130,$^{2}$Friedrich-Alexander Universitat Erlangen-Nurnberg}
\thanks{A. Garson:agarson3@hbar.wustl.edu}}

\maketitle

\begin{abstract}
%\boldmath
The quality of Cadmium Zinc Telluride (CZT) detectors is steadily improving. 
For state of the art detectors, readout noise is thus becoming an increasingly important 
factor for the overall energy resolution. In this contribution, we present measurements 
and calculations of the dark currents and capacitances of 0.5 cm thick CZT detectors 
contacted with a monolithic cathode and 8$\times$8 anode pixels on a surface of 
2$\times$2~cm$^2$. Using the NCI ASIC from Brookhaven National lab as an example, 
we estimate the readout noise caused by the dark currents and capacitances.
Furthermore, we discuss possible additional readout noise caused by 
pixel-pixel and pixel-cathode noise cross-coupling.   
\end{abstract}

% \fi
%\IEEEpeerreviewmaketitle

\section{Introduction}
\thispagestyle{empty}
\IEEEPARstart{T}{here} are multiple applications for the room-temperature semi-conductor Cadmium Zinc Telluride (CZT), ranging from medical imaging over homeland security to astroparticle physics experiments. The high efficiency and good spectral and spatial resolution of CZT make it an attractive material for detecting and measuring photons in the energy range from a few keV to a few MeV.  As the fractional yield of high-quality crystals increases (and the cost is reduced), CZT will become even more prolific in radiation detection systems.  

Limits on the performance of of CZT detector systems depend on characteristics of both the detector 
and the readout electronics.  State-of-the-art CZT detectors combine excellent homogeneity over 
typical volumes between 0.5$\times$2$\times$2 cm$^3$ and 1.5$\times$2$\times$2 cm$^3$ with high 
electron $\mu\tau$-products on the order of $10^{-2}$ cm$^2$ V$^{-1}$. 
As the best thick CZT detectors achieve now 662 keV energy resolutions 
better than 1\% FWHM (full width half maximum), low-noise readout becomes 
increasingly more important. In the following, we will present leakage current and capacitance 
measurements performed on CZT detectors from the company Orbotech Medical Solutions \cite{Orb}. 
Orbotech uses the Modified Horizontal Bridgman process to grow the CZT substrates. The process gives 
substrates with excellent homogeneity, but a somewhat low bulk resistivity of 10$^9$ $\Omega$ cm. 
In earlier work, several groups including ourselves have shown that pixel-cathode dark currents can 
be suppressed efficiently by contacting the substrates with high-work function cathodes \cite{Ira,Jaesub}.
We are currently testing Orbotech detectors with a wide range of thicknesses and with
a range of pixel pitches (see Fig. 1, and Qiang et al., 2007).
In this contribution, we present measurements of the dark currents and capacitances
of an Orbotech CZT detector (0.5$\times$2$\times$2 cm$^3$, 8$\times$8 pixel,2.4mm pitch, 1.6mm pixel side-length), 
and discuss the resulting readout noise. In Sect. 2, the ASIC used as a benchmark for 
noise calculations is described, and the noise model parameters are given. 
The results of dark current and capacitance measurements are described in Sect. 3.
In Sect. 4 the resulting noise is estimated, and in Sect. 5 pixel-pixel 
and pixel-cathode noise cross-coupling is discussed. 
In Sect. 5, the results are summarized.
%\hfill January 11, 2007
\section{Noise Model}
As a reference for our noise calculations, we use the ``NCI ASIC'' developed by Brookhaven 
National Laboratory and the Naval Research Laboratory for the readout of Si strip 
detectors (De Geronimo et al., 2007). 
The self-triggering ASIC comprises 32 channels. Each front-end channel provides 
low-noise charge amplification for pulses of selectable polarity, shaping with 
a stabilized baseline, adjustable discrimination, and peak detection with 
an analog memory. The channels can process events simultaneously, and the read 
out is sparsified. The ASIC requires 5 mW of power per channel. 
 
\begin{figure}[!t]
  \centering
  \includegraphics[width=2.5in]{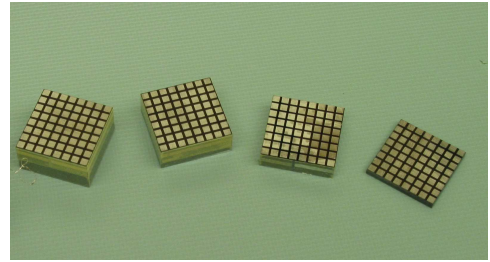}
  
  \caption{Orbotech Cadmium Zinc Telluride (CZT) detectors. From left to right, the detectors 
  have volumes of 1$\times$2$\times$2 cm$^3$, 0.75$\times$2$\times$2 cm$^3$,
  0.5$\times$2$\times$2 cm$^3$, and 0.2$\times$2$\times$2 cm$^3$.}
  \label{CZTs}
\end{figure}
We use the following noise model to calculate the equivalent noise charge (ENC):
\[   ENC^{2} = \left[A_{1}\frac{1}{\tau_{P}}\frac{4kT}{g_{m}}+A_{3}\frac{K_{f}}{C_{G}}\right](C_{G}+C_{D})^{2} \]
\begin{equation}
  +A_{2}\tau_{P}2q(I_{L}+I_{RST})
\end{equation}
where $A_{1},A_{2}$,and $A_{3}$ characterize the pulse shaping filter, 
$\tau_{p}$ is the pulse peaking time, $C_{D}$ and $C_{G}$ are the detector 
and MOSFET capacitances, respectively, $g_{m}$ is the MOSFET transconductance, 
$K_{f}$ is the 1/f noise coefficient, $I_{L}$ is the detector leakage current, 
and $I_{RST}$ is the parallel noise of the reset system (De Geronimo et al.,2002). 
For a given detector ($C_{D},I_{L}$) and ASIC ($A_{1-3},C_{G},g_{m},K_{f}$) 
$\tau_{p}$ can be optimized to reduce the ENC. 
For the NCI ASIC, we use \cite{Ger,Gianluigi-Detailed-Noise-Discussion}: $A_{1}$=0.89, $A_{2}$=$A_{3}$=0.52, $K_{f}$=$10^{-24}$, $C_{G}$=6pF, $g_{m}$=8mS, and $I_{RST}$=50pA.

\section{Dark Current and Capacitance in Orbotech CZT Detectors}
\begin{figure}[!t]
  \centering
  \includegraphics[width=4.0in]{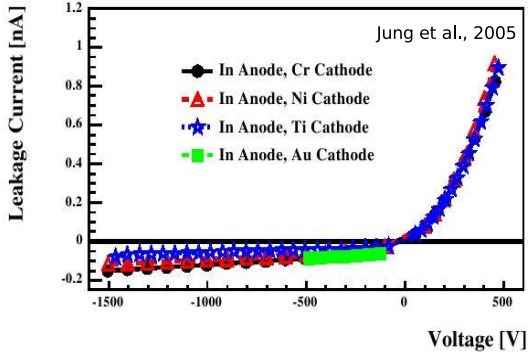}
\vspace*{-1cm}
  \caption{Current-voltage measurements of 0.5cm-thick CZT detectors with various cathode contact materials. 
For all materials, the leakage current is $<$0.2 nA/pixel at biases up to -1500 Volts.}
  \label{IV}
\end{figure}
Figure \ref{IV} shows the IV curves for one 2$\times$2$\times$0.5 cm$^3$ Orbotech CZT detector, 
for different cathode contact materials. The preferred cathode material is Au, as Au cathodes 
give leakage currents $<$0.2 nA/pixel at a cathode bias voltage of -1500 Volts,
and give slightly better spectroscopic performance than other cathode materials.

\begin{figure}[!t]
\vspace*{1cm}
  \centering
  \includegraphics[width=2.0in]{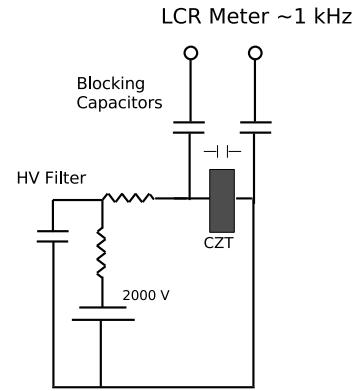}  
\vspace*{0.8cm}
  \caption{Circuit diagram for the pixel-cathode capacitance measurements. Blocking capacitors are used to
		protect the LCR meter from the detector bias voltage. A low pass filter is used to decouple the 
		detector from the high voltage supply at the frequencies used by the LCR meter.
		The set-up makes it possible to measure the capacitance as function of bias voltage. 
		We used cathode biases down to -2000 V.}
  \label{Cap2}
\end{figure}

We used a commercial capacitance meter to measure the capacitance between all pixels and the cathode.
The measurement set-up is shown in Fig. \ref{Cap2}. High voltage blocking capacitors were used to protect 
the LCR meter from the detector bias voltage. A low pass filter 
was used to isolate the LCR-meter-detector circuit from the high voltage supply at the kHz frequencies 
used by the LCR meter. Largely independent of bias voltage, we measure a capacitance of 9 pF for 
all 64 pixels, corresponding to a pixel-cathode capacitance of 0.14 pF per pixel.
The measured result agrees well with a simple estimate of the anode to cathode
capacitance: using a dielectric constant of $\epsilon=$ 10 \cite{Spp}, a parallel plate 
capacitor with the same dimensions as our detector has a capacitance of 8 pF. 

The measurements of the pixel-pixel capacitances resulted in upper limits of $<$1~pF.
For inner pixels (non-border pixels) we estimated the pixel-pixel capacitances with 
the same 3-D Laplace solver that we are using to model the response of CZT detectors 
from our fabrication \cite{Ira}. The code determines the potential inside a 
large grounded box that houses the detector. The capacitance between one pixel and 
its neighbors is determined by setting the voltage of the one pixel to 
$\Delta V=$ 1 V while keeping the other pixels and the cathode at ground potential. 
The charge $\Delta Q$ on the biased pixel and on the neighboring pixels is 
determined with the help of Gauss' law and appropriate Gaussian surfaces.
The procedure gives the capacitances $C=\Delta Q/\Delta V$. We obtain a next-neighbor capacitance
of 0.06 pF, and a diagonal-neighbor capacitance of 0.02 pF.
\section{Equivalent Noise Charge of CZT ASIC Readout Electronics}
\begin{figure}[!t]
  \centering
  \includegraphics[width=3.5in]{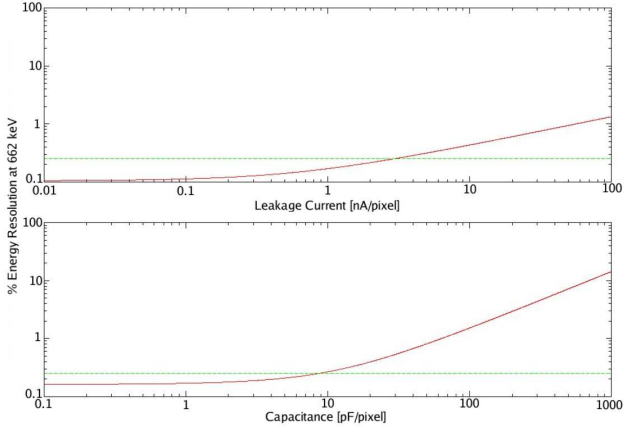}
  \caption{The two panels show the electronic readout noise contribution to the total FWHM energy resolution as a function of detector dark current (top,red line) and detector capacitance (bottom, red line). The calculations are made for the NCI ASIC. A detector capacitance of 1 pF was assumed for the top plot. The electronic noise ($\sim$0.1$\%$ FWHM) is independent of the leakage current $<$0.2 nA/pixel. The bottom figure assumes a leakage current of 1 nA and the resulting electronic noise is constant ($\sim$0.15$\%$ FWHM) for detector capacitances smaller than 2 pF/pixel.}
  \label{results}
\end{figure}

With the previous results, we can now evaluate Equation (1). In the context of 662 keV energy 
depositions (assuming 4.64 eV per electron-hole pair generation, \cite{Spp}),  
Fig.\ \ref{results} plots the FWHM contribution (red line) of the readout electronics's ENC 
as a function of dark current $I_{L}$ (top) and pixel capacitance $C_{D}$ (bottom). 
For the upper plot, $C_{D}$ is held constant at 1.0 pF and for the 
lower plot $I_{L}$ was fixed at 1 nA. In both panels, the green line marks a readout noise contribution 
of 0.25\% FWHM to the 662 keV energy resolution. At the $^{\sim}_<$ 0.25\%-level, the contribution of the 
readout noise to the detector energy resolution is negligible. For the specific ASIC considered here,
we see that leakage currents up to $\sim$3 nA and pixel capacitances $\sim$10 pF are acceptable.
The leads between the readout ASIC and the detector should be sufficiently short 
to stay below 10 pF. 
\section{Pixel-Pixel and Pixel-Cathode Noise Cross-Coupling}
In this section we consider possible additional noise contributions arising from the capacitive 
coupling between adjacent pixels and between a pixel and the cathode. The capacitive coupling can 
result in amplifier noise from one channel being injected into the other channel. In the following 
we use the terminology used by Spieler (2005), and assume that all pixels and the cathode are read 
out by identical preamplifiers. We first consider pixel-pixel noise cross coupling (compare Fig. 5)
and consider the coupling between a pixel, its four nearest neighbors and the cathode.  
The output noise voltage ($\nu{n0}$) of an amplifier creates a noise current, $i_{n}$.
\begin{equation}
  i_{n} = \frac{\nu_{n0}}{\frac{1}{\omega C_{f}}+\frac{1}{\omega \left(4C_{SS}+C_{b}\right)}}
\end{equation}
Here, $C_{SS}$, $C_{B}$, and $C_{f}$ is the pixel-pixel capacitance, the pixel-cathode capacitance, 
and the amplifier capacitance, respectively. The current is divided among the capacitively coupled 
channels in proportion to the coupling capacitance. The fraction of $i_{n}$ going to a nearest neighbor is
\begin{equation}
  \eta_{nn} = \left(4 + \frac{C_{B}}{C_{SS}}\right) \approx \frac{1}{6}.
\end{equation}
Adding the additional noise from the four nearest neighbors in quadrature,  we find 
that the pixel-pixel crosstalk will increase the electronic noise by:
\begin{equation}
  \sqrt{4}\nu_{nn} = 2\frac{\eta_{nn}i_{n}}{\omega C_{f}} \approx 8\%\nu_{n0}
\end{equation}.
In most applications, one reads out the pixels {\it and} the cathode. For single-pixel events, 
the pixel-to-cathode signal ratio can be used to correct the anode signal for the depth of 
the interaction. For multiple-pixel events, the time offset between the cathode signal and the
pixel signals can be used to perform a proper depth of interaction correction for each individual pixel.
The pixel-cathode noise cross-coupling can be more significant. The equivalent noise charge from 
cathode noise being injected into pixels, $Q_{CP}$, depends on the number of pixels ($n_{pix}$) 
and the ratio of the feedback capacitance to the detector capacitance \cite{Spp}:
\begin{equation}
  Q_{CP} = \frac{Q_{n0}}{1 + n_{pix}\frac{C_{d}}{C_{f}}}.
\end{equation}
Here $n_{pix}$=64 is the number of pixel, $C_{d}$= 7 pF is the capacitance between the 
cathode and all the pixels, and $C_{f}$=50 fF is the preamplifier feedback capacitance.
With these values, the cathode noise can increase the readout noise of the anode channels
by 68\%.
\section{Summary}
We measured pixel-cathode dark currents and the pixel-cathode capacitances, both as function of detector bias voltage. 
The measurements give dark currents well below a nA per pixel, and a pixel-cathode capacitance of 0.14 pF per pixel.
\begin{figure}[!t]
  \centering
  \includegraphics[width=3.5in]{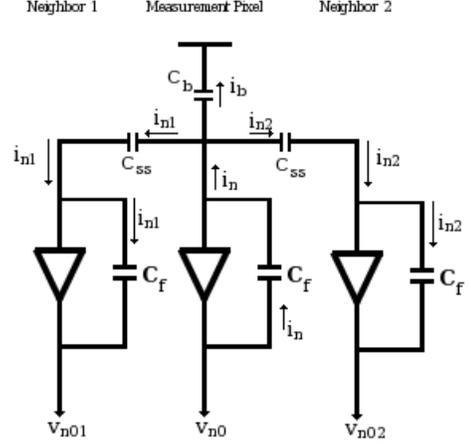}
  \caption{The diagram illustrates the cross-coupling between readout channels of an ASIC. The noise voltage, $\nu_{n0}$, at the output of the measurement pixel's amplifier (center channel) sees an infinite resistance at the amplifier input. The resulting noise current, $i_{n}$, is divided into the cathode and nearest neighbor pixel channels in proportion to their capacitances.   }
  \label{Spieler}
\end{figure}
The pixel-pixel capacitances were smaller than the accuracy of our measurements, and we determined them with the help
of a 3D Laplace solver. We obtain the result that pixel-pixel capacitance is 0.06 pF for direct neighbors and 
0.02 pF for diagonal neighbors. For a state-of-the-art ASIC as the NCI ASIC used as a benchmark here, 
the noise model predicts a very low level of readout noise. With these nominal capacitance values,
pixel-pixel noise cross-coupling is a minor effect, but cathode-pixel noise cross coupling can be significant.
In practice, the readout noise will be higher owing to additional stray capacitances from the detector mounting 
and the readout leads, and from pick-up noise. For the design of a readout system, short leads and a proper 
choice of the detector mounting board substrate are thus of utmost importance. 
\section*{Acknowledgments}
We gratefully acknowledge Gianluigi De Geronimo and Paul O'Connor for information concerning the NCI ASIC. 
This work is supported by NASA under contract NNX07AH37G, and  by the DHS under contract 2007DN077ER0002. 
\ifCLASSOPTIONcaptionsoff
  \newpage
\fi

\end{document}